\documentclass[mathematics,article,accept,pdftex,oneauthor]{Definitions/mdpi}
\usepackage{booktabs}
\usepackage{multirow}
\usepackage{siunitx}
\usepackage{booktabs}
\usepackage[linesnumbered,boxed]{algorithm2e}

\firstpage{1}
\pubvolume{1}
\issuenum{1}
\articlenumber{0}
\pubyear{2026}
\copyrightyear{2026}
\datereceived{ }
\daterevised{ }
\dateaccepted{ }
\datepublished{ }

\makeatletter
\let\@origmaketitle\@maketitle
\renewcommand{\@maketitle}{%
  \let\origincludegraphics\includegraphics
  \renewcommand{\includegraphics}[2][]{%
    \def\@tempa{##2}%
    \@expandtwoargs\in@{.eps}{\@tempa}%
    \ifin@ \else \origincludegraphics[##1]{##2}\fi
  }%
  \@origmaketitle
  \let\includegraphics\origincludegraphics
}
\fancypagestyle{plain}{%
  \fancyhf{}%
  \fancyfoot[C]{\footnotesize\thepage}%
}
\renewcommand{\cright}{}
\renewcommand{\leftcolDates}{}
\renewcommand{\leftcolumnUpdatelogo}{}
\renewcommand{\leftcolumnCorrstatement}{}
\renewcommand{\leftcolumnCright}{}
\makeatother

\newgeometry{left=2.5cm, right=2.5cm, top=1.27cm, bottom=1.08cm,
  includehead, includefoot, footskip=36pt, headsep=24pt}
\fancyheadoffset{0pt}
\fancyfootoffset{0pt}
\newcolumntype{N}{>{\ttfamily\footnotesize}l}
\newcommand{\otm}{\mathrm{otm}}
\newcommand{\erfcx}{\operatorname{erfcx}}
\newcommand{\logonep}{\operatorname{log1p}}
\newcommand{\lfktwenty}{LFK-2026}
\newcommand{\lfktwentyc}{LFK-2026C}

\Title{Explicit Rational Formulae for Bachelier (Normal) Implied Volatility}

\Author{Fabien Le Floc'h\orcidA{}}

\AuthorNames{Fabien Le Floc'h}

\address{%
Independent researcher; fabien@2ipi.com}

\corres{Correspondence: fabien@2ipi.com}

\abstract{%
We present two explicit rational formulae for Bachelier, or normal, implied
volatility.  The formulae take the option price, forward, strike, and expiry as
inputs and return the implied normal volatility without iteration.  They follow
the branch structure of LFK-4, but use the simpler near-the-money variable
given by the absolute forward-strike difference divided by the tail time value,
avoiding a logarithm and a small-argument Taylor branch in that
region.  LFK-2026 is the accuracy-oriented formula and approximates reciprocal
absolute standardized moneyness directly in the far tail.  LFK-2026C keeps the
same shifted out-of-the-money rational tail approximation, but splits the
near-the-money branch into a very small low-
\(u\) rational and a mid-range rational.  In double precision tests both remain
close to machine accuracy, while LFK-2026C is the faster scalar implementation
on the current benchmark mix.}


\begin{document}

\section{Introduction}

The Bachelier model, introduced in Bachelier's 1900 thesis~\cite{bachelier1900},
is widely used when prices are naturally described by a normal volatility rather
than a lognormal volatility.  Its call price is explicit once the volatility is
known, but the inverse
problem---recovering the normal volatility from an observed option price---has no
elementary closed form.  A root finder is robust, but it is often too expensive
for calibration, risk, and large grid calculations where the inverse must be
evaluated millions of times.

This is not only a numerical issue.  In interest-rate derivatives, normal or
Bachelier volatilities are the usual convention when forwards, strikes, or rates
can be close to zero or negative.  Empirical work on low-rate swaption markets
therefore considers Black, Bachelier, and shifted-lognormal volatility quotes
across currencies including EUR, USD, GBP, and JPY~\cite{patel2018quotes}; in
particular, the convention has long been familiar in Japanese-yen interest-rate
markets.  The same modelling issue briefly appeared outside rates in April 2020,
when WTI crude oil futures traded below zero.  CME Clearing then
announced a switch of its options pricing and valuation model to Bachelier to
accommodate negative underlying futures prices and negative strikes
\cite{cme2020bachelier}; Choi, Kwak, Tee, and Wang~\cite{choi2022guide} review
this temporary commodity-exchange transition from Black--Scholes to Bachelier.

In interest-rate markets the same implied quantity is often quoted as
\emph{basis-point volatility}, or bp vol.  This is not a separate volatility
model: it is the normal, equivalently Bachelier, volatility expressed in the
natural absolute units of the underlying.  For a rate forward quoted as a
decimal, a normal volatility \(\sigma\) corresponds to \(10^4\sigma\) basis
points per square-root year.  This contrasts with Black or lognormal volatility,
which is dimensionless and measures proportional rather than absolute moves.

We use the standard option terminology throughout: at the money (ATM) means
\(F=K\); for a call, in the money (ITM) means \(F>K\), and out of the money
(OTM) means \(F<K\).  The formulae are written with the OTM time value, so the
same expressions also cover the reflected ITM side through put--call symmetry.

Explicit approximations fill this gap.  The LFK-4 approximation~\cite{lefloc2016basispoint}
is a standard reference: it transforms the inverse problem into a small number
of rational approximations, using one branch near the money and three branches in
the OTM tail.  Related work includes the direct arithmetic Brownian-motion
implied-volatility approximation of Choi, Kim, and Kwak~\cite{choi2009}.  That
work is useful as a compact closed-form reference, but it is not aimed at the
near-machine-epsilon accuracy sought by LFK-4 and by \lfktwenty{} here.  For
context, J{\"a}ckel's rational inversion of the Black--Scholes formula~\cite{jaeckel2017}
shows the same high-accuracy philosophy in a different model; his normal
implied-volatility approximation is also a popular production reference.
QuantLib~\cite{quantlib} includes an implementation in this family, but it is
not a fully faithful transcription of J{\"a}ckel's original code and is less
accurate.  The timings below therefore compare the local Java implementations
used in this study rather than the QuantLib port.  The goal here is
to keep the practical LFK-4 structure while improving the accuracy--cost
trade-off.  The two formulae below are designed to be portable:
they use only arithmetic operations and a logarithm in the OTM routing variable.
Both \lfktwenty{} and \lfktwentyc{} avoid the final OTM square root by fitting
\(1/|d|\) directly.

Accurate validation of these formulae requires an accurate forward Bachelier
price.  Direct formulas for the OTM time value suffer cancellation in the tails,
which can make an inverse approximation look worse than it is.  For benchmarking
we therefore compute the OTM time value with a symmetric \(\erfcx\) expression;
this makes the comparison primarily about the inverse formula rather than about
loss of precision in the price used as input.

\section{Setup and Notation}

Let
\[
  x = F-K, \qquad v = \sigma\sqrt{T}, \qquad d = \frac{x}{v}.
\]
The undiscounted Bachelier call price is
\[
  C = x\Phi(d)+v\phi(d).
\]
Thus ATM corresponds to \(m=|x|=0\).  Away from ATM, the call is ITM for
\(x>0\) and OTM for \(x<0\); the reflected put side gives the same time value.
As in LFK-4, the inverse is expressed in terms of the out-of-the-money time value
and the absolute moneyness,
\[
  c_{\otm} =
  \begin{cases}
    C-x, & x>0,\\
    C, & x\le 0,
  \end{cases}
  \qquad m=|x|.
\]
At the money, where \(m=0\), the inverse is exact:
\[
  \sigma = \frac{C\sqrt{2\pi}}{\sqrt{T}}.
\]
For \(m>0\), define
\[
  g=\frac{c_{\otm}}{m}, \qquad \alpha=0.15,
  \qquad \beta_s=-\ln\alpha, \qquad \beta_e=300\ln(10).
\]
The ITM/near-ATM branch is used when \(g>\alpha\).  The OTM branch uses
\[
  \tilde\eta = -\frac{\ln g+\beta_s}{\beta_e-\beta_s}.
\]
The three OTM zones are
\[
\begin{array}{c|c}
j & \tilde\eta\text{ range}\\
\hline
1 & [0,0.011)\\
2 & [0.011,0.105)\\
3 & [0.105,1].
\end{array}
\]

For all coefficient tables, rational functions are written in ascending powers:
\[
  R_{p,q}(y;a,b)
  = \frac{P_p(y)}{Q_q(y)}, \qquad
  P_p(y)=\sum_{i=0}^{p}a_i y^i, \qquad
  Q_q(y)=\sum_{i=0}^{q}b_i y^i, \quad b_0=1.
\]
For numerical evaluation, the same polynomials may be evaluated by ordinary
Horner steps in descending powers.

\section{Reference: LFK-4}

LFK-4~\cite{lefloc2016basispoint} uses
\[
  z=\frac{m}{m+c_{\otm}}, \qquad
  \eta=\frac{-z}{\ln(1-z)}.
\]
For small \(z\), the original algorithm evaluates \(\eta\) by a Taylor expansion;
otherwise it uses \(\logonep(-z)\).  This avoids cancellation in the near-ATM
limit.  For \(g>\alpha\), LFK-4 returns
\[
  \sigma = \frac{m+c_{\otm}}{\sqrt T}\,h_{\mathrm{LFK}}(\eta),
\]
with a P7/Q5 rational.  For \(g\le\alpha\), it fits \(d^2\) with three P9/Q7
rationals in \(\tilde\eta\), using the zones listed above.

\section{\lfktwenty{}}

\lfktwenty{} uses the no-log ITM variable
\[
  u=\frac{z}{1-z}=\frac{m}{c_{\otm}}.
\]
For \(g>\alpha\), it uses a P10/Q9 rational:
\[
  \sigma = \frac{m+c_{\otm}}{\sqrt T}\,R_{10,9}(u;a^{26}_I,b^{26}_I).
\]
In the OTM branch, \lfktwenty{} approximates \(1/|d|\) directly:
\[
  W^{26}_j(\tilde\eta)=R_{10,9}(\tilde\eta;a^{26}_j,b^{26}_j),
  \qquad
  \sigma = \frac{m}{\sqrt T}\,W^{26}_j(\tilde\eta).
\]
This removes the final square root in the OTM branch.  The method uses more
coefficients in the ITM branch than \lfktwentyc{}, but the two methods share the
same OTM rational approximation and the same direct OTM accuracy envelope.

\section{\lfktwentyc{}}

\lfktwentyc{} uses the same no-log ITM variable \(u=m/c_{\otm}\), but splits the
ITM branch at \(u=0.20\).  For small \(u\) it uses a compact P4/Q4 rational,
and otherwise it uses a P9/Q8 rational:
\[
  \sigma = \frac{m+c_{\otm}}{\sqrt T}\,
  \begin{cases}
    R_{4,4}(u;a^{26C}_L,b^{26C}_L), & u<0.20,\\
    R_{9,8}(u;a^{26C}_I,b^{26C}_I), & u\ge 0.20.
  \end{cases}
\]
In the OTM branch it uses the same shifted-zone \(1/|d|\) approximation as
\lfktwenty{}:
\[
  W^{26C}_j(\tilde\eta)=R_{10,9}(\tilde\eta;a^{26}_j,b^{26}_j),
  \qquad
  \sigma = \frac{m}{\sqrt T}\,W^{26C}_j(\tilde\eta).
\]
This is the fast variant.  It keeps the OTM accuracy of \lfktwenty{} while
reducing the average cost of the ITM/near-ATM branch.

\section{Numerical Results}
\subsection{Implementations of the Bachelier Formula}
A naive implementation of the Bachelier formula, used in many production systems consists in using the standard routines for the cumulative normal distribution and normal probability density function.
This leads to somewhat large errors in the prices of very out-of-the-money options. A nice way to look at it is to first compute the prices considering a fixed volatility $\sigma = 1$, time-to-maturity $T=1$, varying $|d|$, and 
to invert the resulting option price back using a bisection method to maximum accuracy. In the bisection method, the reference price is given by a multiple precision arithmetic implementation of the Bachelier formula.

\begin{figure}[ht]
\centering
\includegraphics[width=\linewidth]{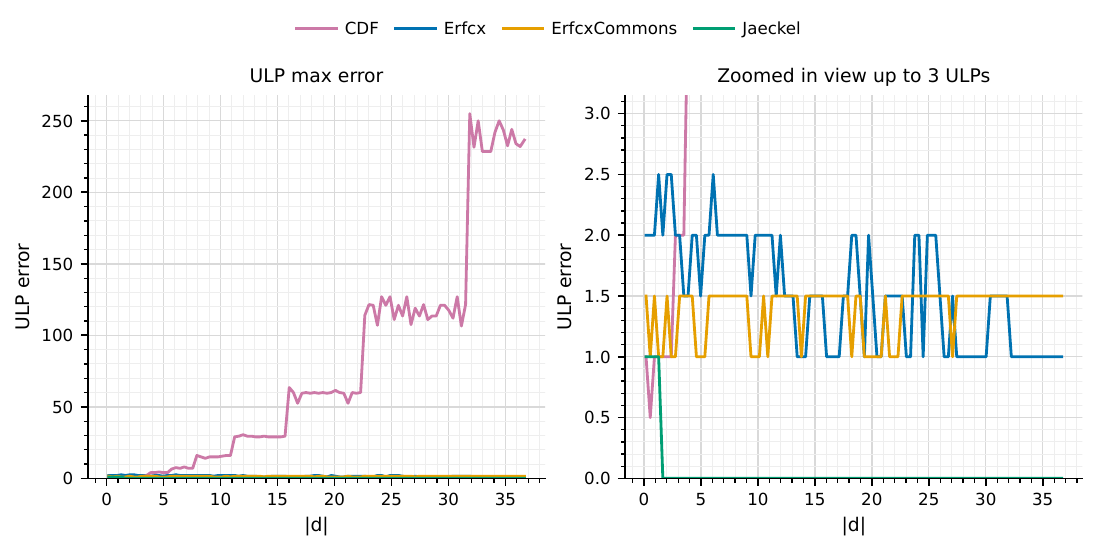}
\caption{Error in the volatility implied from different Bachelier option price implementations.}
\label{fig:julia_bins_pricevols}
\end{figure}

In Figure \ref{fig:julia_bins_pricevols}, we sample 100,000 uniformly distributed values $|d| \in [0,35]$ and compute the maximum relative error in ULP (Unit in the Last Place) for the Bachelier implied volatility in 100 uniform buckets. 
This clearly shows the loss of accuracy in the very out-of-the-money region for the naive implementation.

A better idea presented in \cite{lefloc2016basispoint} is to use a symmetric \(\erfcx\) expression for the OTM time value.  With
\(a=m/v\),
\[
  c_{\otm}=e^{-a^2/2}\left(\frac{v}{\sqrt{2\pi}}
  -\frac{m}{2}\erfcx\left(\frac{a}{\sqrt 2}\right)\right),
\]
which avoids cancellation in both deep ITM and deep OTM cases.
There are different implementations of the erfcx function, we show the default Julia SpecialFunctions.jl implementation, based on Steven G. Johnson Faddeeva package, and the more accurate implementation found in the Java Apache Commons Numbers library.
It is closer to the money ($|d| \leq 5$) that the latter results in a (small) improvement in accuracy. Cody \cite{cody1993algorithm} also provides an accurate representation of erfcx, and its most accurate pieces are reused in Apache Commons Numbers.

Peter J\"ackel \cite{jaeckel2017} gives an even more accurate implementation of the Bachelier formula. Inspired by Cody's erfcx formula, he derives a rational function representation of 
\begin{equation*}
  \tilde{\Phi}(u) = \Phi(u) + \frac{\phi(u)}{u},
\end{equation*}
for $u \in \mathbb{R}$. This is then used to compute the option price through $C = x\tilde{\Phi}(d)$. Except very close to the money, the ULP error in implied volatility is essentially zero, which is remarkable.

Noticing that the naive Bachelier formula is very accurate for $|d| < d_{\textmd{J}}$, with $d_{\textmd{J}} = 0.61200318096248076056$ being the first cutoff of $\tilde{\Phi}$ implementation, we may combine J\"ackel's representation for
$|d| > d_{\textmd{J}}$ with the naive Bachelier formula for $|d| \le d_{\textmd{J}}$. This has the dual benefit of being slightly more accurate (although we may really wonder if it makes sense to care so much about errors below 1 ULP) and using the classic naive implementation for $|d|$ values which correspond to typical real market prices. We will however consider the J\"ackel formula to compute reference prices in 64-bit floating point numbers, unless otherwise stated.

\subsection{Accuracy of the Rational Formulae for the Bachelier Implied Volatility}

We start by reproducing the example of Le Floc'h \cite{lefloc2016basispoint} of options of maturity $T=1$ with forward $F=1.0$ and  volatility $\sigma=1.0$ with varying strikes in Table \ref{tab:lefloch1}. However, this kind of test can be misleading as there are not enough strikes sampled to hit the worst case scenario of a given approximation. We added a strike at $2.7838494978971955$ to show a worst case for LFK-4.
\begin{table}[ht]
  \centering
  \caption{Error in the Bachelier volatility implied from J\"ackel Bachelier price implementation for $\sigma = 1.0, T = 1.0, F = 1.0$.}
  \label{tab:lefloch1}
  \begin{tabular}{l r r r r}
    \toprule
    Strike & LFK-4 & LFK-2026 & LFK-2026C & Jaeckel2017\\\midrule
    1.000010 & 4.44e-16 & 2.22e-16 & 1.11e-16  & 0\\
    1.006660 & 0 & 2.22e-16 & 2.22e-16 &  0\\
    2.000000 & 2.22e-16 & 0 & 0 &  0\\
    2.783849 & 1.33e-15 & 0 & 0 &  0\\
    4.000000 & 4.44e-16 & 0 & 0 &  0\\
    8.800000 & 0 & 2.22e-16 & 2.22e-16 & 0\\
    9.000000 & 0 & 4.44e-16 & 4.44e-16 &  0\\
    30.00000 & 2.22e-16 & 4.44e-16 & 4.44e-16 & 0\\\bottomrule
  \end{tabular}
\end{table}

It is a better idea to randomly sample from a space of parameters that covers the whole
region of interest for the argument $d = (F-K)/(\sigma\sqrt{T})$.  We do
this in Table \ref{tab:bucket-errors}, where we sample $40,000\,000$ uniform
values in several buckets for $|d| \in [0,35]$ and compute the maximum relative error in ULP for the
Bachelier implied volatility.

\begin{table}[htbp]
\centering
\caption{Absolute volatility error statistics per bucket and overall}
\label{tab:bucket-errors}
\small
\setlength{\tabcolsep}{4pt}
\begin{tabular}{llrrrrr}
\toprule
Bucket & Approximation & Cases & Max & Worst $d$ & $p_{95}$ & $p_{99}$ \\
\midrule

\multicolumn{7}{l}{\textbf{Bucket 0.00 --- range [0.00, 1.00]}} \\
0.00 & LFK-4       & 40,000,000 & 1.33e-15 & 0.5306090208459261 & 4.44e-16 & 6.66e-16 \\
0.00 & LFK-2026    & 40,000,000 & 1.22e-15 & 0.7783018038945502 & 4.44e-16 & 4.44e-16 \\
0.00 & LFK-2026C   & 40,000,000 & 9.99e-16 & 0.7700211092071692 & 3.33e-16 & 4.44e-16 \\
0.00 & Jaeckel2017 & 40,000,000 & 5.55e-16 & 0.6121702394114410 & 2.22e-16 & 2.22e-16 \\
\midrule

\multicolumn{7}{l}{\textbf{Bucket 1.00 --- range [1.00, 2.00]}} \\
1.00 & LFK-4       & 40,000,000 & 1.33e-15 & 1.7896497885619913 & 8.88e-16 & 9.99e-16 \\
1.00 & LFK-2026    & 40,000,000 & 8.88e-16 & 1.9791769043189662 & 3.33e-16 & 4.44e-16 \\
1.00 & LFK-2026C   & 40,000,000 & 8.88e-16 & 1.9791769043189662 & 3.33e-16 & 4.44e-16 \\
1.00 & Jaeckel2017 & 40,000,000 & 4.44e-16 & 1.1742267194679015 & 2.22e-16 & 2.22e-16 \\
\midrule

\multicolumn{7}{l}{\textbf{Bucket 2.00 --- range [2.00, 32.00]}} \\
2.00 & LFK-4       & 40,000,000 & 1.55e-15 & 2.6491128666595603 & 6.66e-16 & 8.88e-16 \\
2.00 & LFK-2026    & 40,000,000 & 1.22e-15 & 4.1572426607873700 & 4.44e-16 & 6.66e-16 \\
2.00 & LFK-2026C   & 40,000,000 & 1.22e-15 & 4.1572426607873700 & 4.44e-16 & 6.66e-16 \\
2.00 & Jaeckel2017 & 40,000,000 & 2.22e-16 & 3.0937788117606120 & 0.00e+00 & 2.22e-16 \\
\midrule

\multicolumn{7}{l}{\textbf{Overall (all buckets combined)}} \\
-- & LFK-4       & 158,816,168 & 1.55e-15 & 2.649113 & 6.66e-16 & 8.88e-16 \\
-- & LFK-2026    & 158,816,168 & 1.22e-15 & 0.778302 & 4.44e-16 & 4.44e-16 \\
-- & LFK-2026C   & 158,816,168 & 1.22e-15 & 4.157243 & 4.44e-16 & 4.44e-16 \\
-- & Jaeckel2017 & 160,000,000 & 5.55e-16 & 0.612170 & 2.22e-16 & 2.22e-16 \\
\bottomrule
\end{tabular}
\end{table}

\section{Pushing Accuracy Further}
\begin{figure}[ht]
\centering
\includegraphics[width=\linewidth]{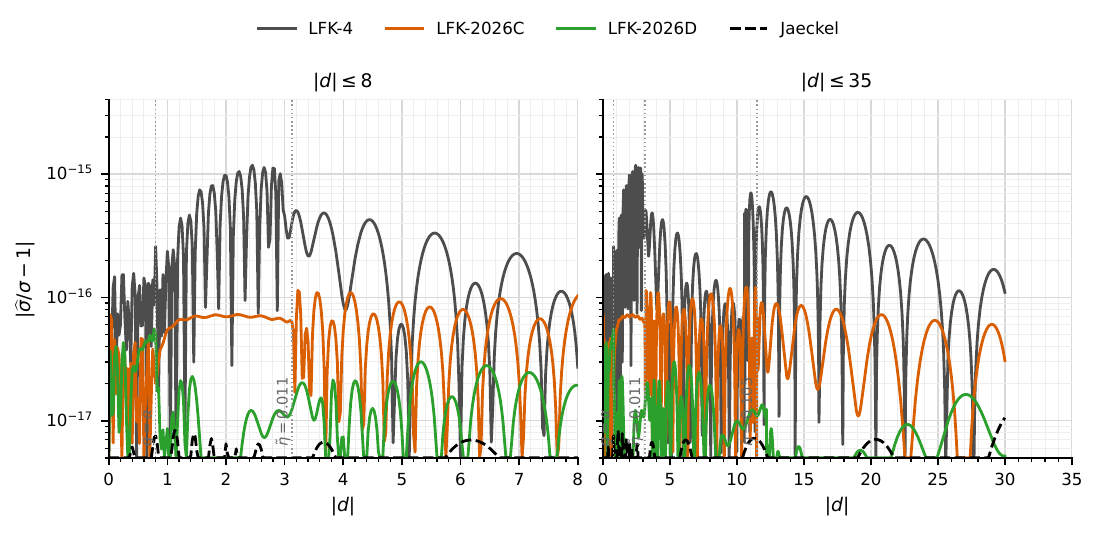}
\caption{Direct BigFloat relative implied-volatility error for LFK-4,
\lfktwenty{}, and \lfktwentyc{}.  The left panel focuses on the central region
\(|d|\le8\), while the right panel shows the full direct grid \(|d|\le35\).  The
dotted vertical lines mark the route boundary \(g=\alpha\) and the two OTM zone
boundaries.}
\label{fig:accuracy-profiles-bigfloat}
\end{figure}

When we look at the error in multiple precision arithmetic, we notice that the new variants LFK-2026 and LFK-2026C are more accurate than LFK-4 (Figure \ref{fig:accuracy-profiles-bigfloat}). We also derived an even more accurate approximation in multiple precision arithmetic called LFK-2026D (we give the coefficients in Appendix \ref{app:LFK2026D-coeffs}).As shown in Table~\ref{tab:bucket-errors}, this does not lead to improved accuracy in 64-bit floating-point arithmetic when using a straightforward Horner-method implementation (Algorithm \ref{algo:horner}) for evaluating the rational functions.
\begin{algorithm}
\caption{Classic Horner scheme for polynomial evaluation \label{algo:horner}}
\KwIn{$x \in \mathbb{R}$, coefficients $c = (c_0, c_1, \dots, c_{n-1})$}
\KwOut{$y = c_0 + c_1 x + c_2 x^2 + \dots + c_{n-1} x^{n-1}$}
$y \gets c_0$\;
\For{$i \gets 1$ \KwTo $n-1$}{
    $y \gets y \cdot x + c_i$\;
}
\Return{$y$}\;
\end{algorithm}
The classic implementation of Horner's method is suboptimal in terms of floating point error. One slight improvement consists in using fused-multiply-add (FMA) instructions in the Horner evaluation. Another advantage of FMA is improved speed (at least on x86 architectures) as shown in Algorithm \ref{algo:fma_horner}.
\begin{algorithm}
\caption{Horner scheme with fused multiply–add (FMA) \label{algo:fma_horner}}
\KwIn{$x \in \mathbb{R}$, coefficients $c = (c_0, c_1, \dots, c_{n-1})$}
\KwOut{$y = \sum_{i=0}^{n-1} c_i x^i$ evaluated with FMA for each step}
$y \gets c_0$\;
\For{$i \gets 1$ \KwTo $n-1$}{
    $y \gets \operatorname{fma}(y, x, c_i)$\;
}
\Return{$y$}\;
\end{algorithm}

If the ultimate goal is to increase accuracy even closer to machine epsilon, one solution is to use the compensated Horner FMA scheme, where the error is accumulated and added back to the main approximation \cite{gll2005}:
\begin{algorithm}
\caption{Compensated Horner scheme with FMA \label{algo:compensated-horner}}
\KwIn{$x \in \mathbb{R}$, coefficients $c = (c_0, c_1, \dots, c_{n-1})$}
\KwOut{$y = \sum_{i=0}^{n-1} c_i x^i$ evaluated with compensated arithmetic using FMA}
$s \gets c_0$\;
$e \gets 0$\;
\For{$i \gets 1$ \KwTo $n-1$}{
    $p \gets s \cdot x$\;
    $\mathit{pe} \gets \operatorname{fma}(s, x, -p)$\;

    $t \gets p + c_i$\;
    $z \gets t - p$\;
    $\mathit{te} \gets (p - (t - z)) + (c_i - z)$\;

    $s \gets t$\;
    $e \gets \operatorname{fma}(e, x, \mathit{pe} + \mathit{te})$\;
}
\Return{$s + e$}\;
\end{algorithm}

The FMA implementation (suffix + in Table \ref{tab:bucket-errors-lfk2026}) is slightly more accurate (by around 2 machine epsilons), and the compensated horner (suffix ++) let LFK-2026++ and LFK-2026C++ reach an maximum error of 2 machine epsilons (4.44e-16).
\begin{table}[htbp]
\centering
\caption{Absolute volatility error statistics per bucket and overall for LFK-2026 variants. The suffix + denotes FMA implementation and ++ denotes the compensated Horner FMA scheme.}
\label{tab:bucket-errors-lfk2026}
\small
\setlength{\tabcolsep}{4pt}
\begin{tabular}{llrrrrr}
\toprule
Bucket & Approximation & Cases & Max & Worst $d$ & $p_{95}$ & $p_{99}$ \\
\midrule

\multicolumn{7}{l}{\textbf{Bucket 0.00 --- range [0.00, 1.00]}} \\
0.00 & LFK-2026+   & 40,000,000 & 8.88e-16 & 0.7631171052449106 & 2.22e-16 & 4.44e-16 \\
0.00 & LFK-2026++  & 40,000,000 & 6.66e-16 & 0.7749365794558335 & 2.22e-16 & 2.22e-16 \\
0.00 & LFK-2026C+  & 40,000,000 & 8.88e-16 & 0.7582758869622748 & 2.22e-16 & 4.44e-16 \\
0.00 & LFK-2026C++ & 40,000,000 & 5.55e-16 & 0.9830956928512565 & 2.22e-16 & 2.22e-16 \\
0.00 & LFK-2026D+  & 40,000,000 & 9.99e-16 & 0.7280022477431374 & 2.22e-16 & 4.44e-16 \\
0.00 & LFK-2026D++ & 40,000,000 & 6.66e-16 & 0.7675775666172642 & 2.22e-16 & 2.22e-16 \\
\midrule

\multicolumn{7}{l}{\textbf{Bucket 1.00 --- range [1.00, 2.00]}} \\
1.00 & LFK-2026+   & 40,000,000 & 6.66e-16 & 1.8237757808060500 & 2.22e-16 & 3.33e-16 \\
1.00 & LFK-2026++  & 40,000,000 & 5.55e-16 & 1.0475809080916290 & 2.22e-16 & 2.22e-16 \\
1.00 & LFK-2026C+  & 40,000,000 & 6.66e-16 & 1.8237757808060500 & 2.22e-16 & 3.33e-16 \\
1.00 & LFK-2026C++ & 40,000,000 & 5.55e-16 & 1.0475809080916290 & 2.22e-16 & 2.22e-16 \\
1.00 & LFK-2026D+  & 40,000,000 & 6.66e-16 & 1.7722558130136150 & 2.22e-16 & 4.44e-16 \\
1.00 & LFK-2026D++ & 40,000,000 & 4.44e-16 & 1.7803130699297630 & 2.22e-16 & 2.22e-16 \\
\midrule

\multicolumn{7}{l}{\textbf{Bucket 2.00 --- range [2.00, 32.00]}} \\
2.00 & LFK-2026+   & 40,000,000 & 1.11e-15 & 9.4045676549501300 & 4.44e-16 & 4.44e-16 \\
2.00 & LFK-2026++  & 40,000,000 & 6.66e-16 & 7.9895611392130190 & 2.22e-16 & 2.22e-16 \\
2.00 & LFK-2026C+  & 40,000,000 & 1.11e-15 & 9.4045676549501300 & 4.44e-16 & 4.44e-16 \\
2.00 & LFK-2026C++ & 40,000,000 & 6.66e-16 & 7.9895611392130190 & 2.22e-16 & 2.22e-16 \\
2.00 & LFK-2026D+  & 40,000,000 & 8.88e-16 & 27.3829069373139650 & 3.33e-16 & 4.44e-16 \\
2.00 & LFK-2026D++ & 40,000,000 & 4.44e-16 & 7.8976758999678350 & 2.22e-16 & 2.22e-16 \\
\midrule

\multicolumn{7}{l}{\textbf{Overall (all buckets combined)}} \\
-- & LFK-2026+   & 158,816,168 & 1.11e-15 & 9.404568 & 3.33e-16 & 4.44e-16 \\
-- & LFK-2026++  & 158,816,168 & 6.66e-16 & 0.774937 & 2.22e-16 & 2.22e-16 \\
-- & LFK-2026C+  & 158,816,168 & 1.11e-15 & 9.404568 & 3.33e-16 & 4.44e-16 \\
-- & LFK-2026C++ & 158,816,168 & 6.66e-16 & 7.989561 & 2.22e-16 & 2.22e-16 \\
-- & LFK-2026D+  & 158,816,168 & 9.99e-16 & 0.728002 & 2.22e-16 & 4.44e-16 \\
-- & LFK-2026D++ & 158,816,168 & 6.66e-16 & 0.767578 & 2.22e-16 & 2.22e-16 \\
\bottomrule
\end{tabular}
\end{table}
The LFK-2026D++ is not more accurate in float64 arithmetic despite being more accurate in multiple-precision arithmetic.

\begin{figure}[ht]
\centering
\includegraphics[width=\linewidth]{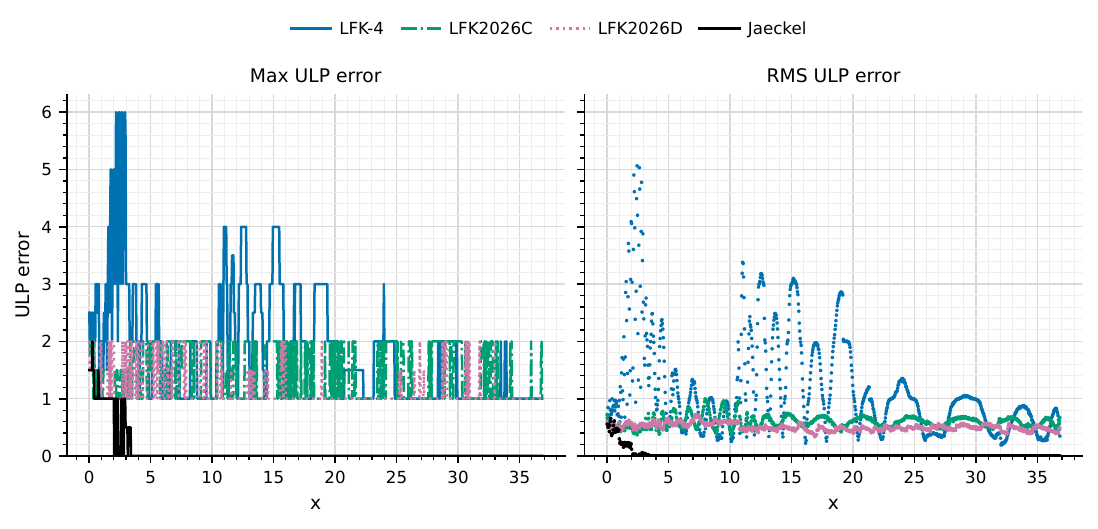}
\caption{Relative implied-volatility error in ULP. LFK-2026C and LFK-2026D use the compensated Horner FMA scheme.}
\label{fig:accuracy-profiles-float64j}
\end{figure}

In Figure \ref{fig:accuracy-profiles-float64j}, we sample the space of normalized OTM Bachelier distances $|d|$ in the range $0 \le |d| \le 35$, uniformly.  For each input, the target volatility is $1.0$, while each implementation returns a 64-bit floating point approximation $y = \text{float64}(\text{approximation}(x))$.  The ULP error is then computed as
\[
\text{ulp}(x) = \frac{|y - y_\text{ref}|}{\text{ulp}(y_\text{ref})},
\]
where $\text{ulp}(y_\text{ref}) = | \text{nextfloat}(y_\text{ref}) - y_\text{ref} |$ is the distance between adjacent 64-bit floating-point numbers at $y_\text{ref}$.  This normalizes the absolute error into units of machine epsilon, so that $\text{ulp} = 1$ corresponds to an error of one floating-point step.  The $|d|$ range is split into 1000 equal-width bins, and for each bin we record the maximum and RMS ULP error over $10^6$ random samples.  This removes call-price rounding and intrinsic-subtraction effects, so the plot shows the rational-approximation ULP error itself.

As a final test, we consider the example from Cui et al. \cite{cui2025tighter}, which consists of a three-dimensional data set of size $40 \times 40 \times 40$ with forward $F=1.0$. Discrete points of $K$ are uniformly generated from $1.01$ to $10$, $\sigma$ from $0.01$ to $0.99$, and $T$ from $0.01$ to $2$. Samples with option prices less than $10^{-20}$ are filtered out, leaving $29\,330$ retained samples.

In Table \ref{tab:3d-grid-errors}, we report a much higher accuracy for J\"ackel algorithm than the authors. The discrepancy is likely the choice of implementation for the Bachelier option price. We use the highly accurate J\"ackel Bachelier option price implementation, while they likely use the naive CDF implementation.
\begin{table}[ht]
\centering
\caption{Absolute volatility error statistics and timing for the 3D grid experiment (Cui et al.)}
\label{tab:3d-grid-errors}
\begin{tabular}{l r r r r r}
\toprule
\textbf{Method} & \textbf{AbsErrorMean} & \textbf{AbsErrorDev} & \textbf{AbsErrorMax} & \textbf{AbsErrorMin}  & \textbf{Time(sec)} \\
\midrule
LFK-2026+    & 1.00e-16 & 9.87e-17 & 6.66e-16 & 0 & 2.30e-08 \\
LFK-2026++   & 7.42e-17 & 7.88e-17 & 5.55e-16 & 0 & 5.13e-08 \\
LFK-2026C+   & 9.97e-17 & 9.87e-17 & 6.66e-16 & 0 & 2.34e-08 \\
LFK-2026C++  & 7.40e-17 & 7.89e-17 & 5.55e-16 & 0 & 5.04e-08 \\
LFK-2026D+   & 9.22e-17 & 9.25e-17 & 6.66e-16 & 0 & 2.47e-08 \\
LFK-2026D++  & 6.88e-17 & 7.50e-17 & 4.44e-16 & 0 & 5.11e-08 \\
J\"ackel-2017  & 3.96e-17 & 5.53e-17 & 3.33e-16 & 0 & 6.58e-08 \\
\bottomrule
\end{tabular}
\end{table}
The LFK-2026+ is nearly three times faster than J\"ackel-2017 and gives up very little in accuracy. LFK-2026D++ is slightly more accurate that the other LFK-2026 formulas on this example, this is consistent with the lower RMSE observed in Figure \ref{fig:accuracy-profiles-float64j}.

\subsection{Timing}
The timing table below uses a rotated-order scalar benchmark and reports medians
over fifty rounds.  This is more stable than the fixed-order main report, whose
single sequential timing slices can move when CPU frequency or scheduling changes
during the run.  Absolute timings are machine- and power-state-dependent; the
relative ordering is the more reproducible part of the table.

\begin{table}[ht]
\centering
\caption{Timing results on an Intel(R) Core(TM) Ultra 5 245KF CPU running Java 25.0.3 for implying the volatility of options of strikes $K \in 	\{ 1.001, 1.01, 1.1, 1.5, 2.0, 3.0, 5.0, 7.0, 8.0,
				0.999, 0.99, 0.9, 0.5, 0.0, -0.5, -1.0 \}$ and forward $F=1$, volatility $\sigma=1$, maturity $T=1$.}
\label{tab:timing}
\begin{tabular}{lrr}
\toprule
Approximation & median ns/call & Relative to LFK-4\\
\midrule
LFK-4      & 12.36 & 1.00\\
\lfktwenty{}      &  9.51 & 0.77\\
\lfktwenty{}+     &  7.85 & 0.64\\
\lfktwenty{}++    & 22.41 & 1.81\\
\lfktwentyc{}     &  9.41 & 0.76\\
\lfktwentyc{}+    &  7.77 & 0.63\\
\lfktwentyc{}++   & 20.50 & 1.66\\
J{\"a}ckel & 25.75 & 2.08\\
\bottomrule
\end{tabular}
\end{table}

\appendix
\section{\lfktwenty{} Coefficients}

\subsection{ITM/near-ATM branch}

\begin{center}
\begin{tabular}{rNN}
\toprule
\(i\) & \(a^{26}_{I,i}\) & \(b^{26}_{I,i}\)\\
\midrule
0  &  2.50662827463100069e+00 & 1.00000000000000000e+00\\
1  &  8.26914966237441540e+00 & 3.79891342328838499e+00\\
2  &  1.10083895644891214e+01 & 5.87074621959482457e+00\\
3  &  7.62695942399991100e+00 & 4.76157470081986212e+00\\
4  &  2.96457253391146036e+00 & 2.18581925252758946e+00\\
5  &  6.51812439800918964e-01 & 5.72905243689549204e-01\\
6  &  7.81257091685455263e-02 & 8.27294623127229206e-02\\
7  &  4.67776338554095131e-03 & 6.05262799121958784e-03\\
8  &  1.17427118047196583e-04 & 1.90358881940080979e-04\\
9  &  8.06959950534549627e-07 & 1.76070970713414280e-06\\
10 & -3.39451638335349906e-10 & \\
\bottomrule
\end{tabular}
\end{center}

\subsection{OTM branch, zone 1}

\begin{center}
\begin{tabular}{rNN}
\toprule
\(i\) & \(a^{26}_{1,i}\) & \(b^{26}_{1,i}\)\\
\midrule
0  &  1.24910559446641112e+00 & 1.00000000000000000e+00\\
1  &  8.30013684602045146e+02 & 9.50178311644246946e+02\\
2  &  2.89118707775471907e+05 & 4.30671117130989209e+05\\
3  &  6.20616261157058701e+07 & 1.19319078802154481e+08\\
4  &  8.81948242946073532e+09 & 2.18139815883858109e+10\\
5  &  8.14953597568851318e+11 & 2.66113331187980811e+12\\
6  &  4.61473754119971406e+13 & 2.08194332615256938e+14\\
7  &  1.32385751716178975e+15 & 9.31612904921204000e+15\\
8  &  1.35683441709766740e+16 & 1.78457095251951296e+17\\
9  &  2.23602102096865640e+16 & 9.01018167981477888e+17\\
10 & -1.15188057059215700e+16 & \\
\bottomrule
\end{tabular}
\end{center}

\subsection{OTM branch, zone 2}

\begin{center}
\begin{tabular}{rNN}
\toprule
\(i\) & \(a^{26}_{2,i}\) & \(b^{26}_{2,i}\)\\
\midrule
0  &  1.25087969033276813e+00 & 1.00000000000000000e+00\\
1  &  2.09344504116440078e+02 & 4.55619235679538008e+02\\
2  & -1.74963499510044603e+04 & 4.19868074702521844e+04\\
3  & -1.28046460022976995e+07 & -1.48653150107890852e+07\\
4  & -1.43888265763526964e+09 & -4.63359666129231930e+09\\
5  & -5.70761754440745773e+10 & -3.39232384615881042e+11\\
6  & -8.98414557472468994e+11 & -8.93831987833861328e+12\\
7  & -5.46607883881420703e+12 & -9.10822517732905625e+13\\
8  & -1.07736581582367109e+13 & -3.32157260465782750e+14\\
9  & -3.54669640321780615e+12 & -3.16204789033010312e+14\\
10 &  4.53512764886659485e+11 & \\
\bottomrule
\end{tabular}
\end{center}

\subsection{OTM branch, zone 3}

\begin{center}
\begin{tabular}{rNN}
\toprule
\(i\) & \(a^{26}_{3,i}\) & \(b^{26}_{3,i}\)\\
\midrule
0  &  1.61713874667576762e+00 & 1.00000000000000000e+00\\
1  &  1.63839367097254751e+02 & 6.08432004257079598e+02\\
2  &  4.52092618390189637e+03 & 3.37041720211591382e+04\\
3  &  4.73452995425928821e+04 & 5.83977840443553636e+05\\
4  &  2.11536138072810922e+05 & 4.03973802769196732e+06\\
5  &  4.14000232322855853e+05 & 1.20766363495907933e+07\\
6  &  3.38490556724923430e+05 & 1.55296990844987314e+07\\
7  &  9.82243936325080576e+04 & 7.83292053799574263e+06\\
8  &  5.74901275345015438e+03 & 1.16810498733679834e+06\\
9  & -1.25257952774401772e+02 & 5.54442814599942540e+03\\
10 &  4.62144628906322019e+00 & \\
\bottomrule
\end{tabular}
\end{center}

\section{\lfktwentyc{} Coefficients}

\subsection{Low-\(u\) ITM/near-ATM branch, \(u<0.20\)}

\begin{center}
\begin{tabular}{rNN}
\toprule
\(i\) & \(a^{26C}_{L,i}\) & \(b^{26C}_{L,i}\)\\
\midrule
0 &  2.50662827463100069e+00 & 1.00000000000000000e+00\\
1 &  4.33803460770355898e+00 & 2.23062541885758314e+00\\
2 &  2.36504153278287266e+00 & 1.63840524330855919e+00\\
3 &  4.35189695439266722e-01 & 4.35646805136993498e-01\\
4 &  1.77075775698172025e-02 & 2.96543644408861981e-02\\
\bottomrule
\end{tabular}
\end{center}

\subsection{High-\(u\) ITM/near-ATM branch, \(u\ge0.20\)}

\begin{center}
\begin{tabular}{rNN}
\toprule
\(i\) & \(a^{26C}_{I,i}\) & \(b^{26C}_{I,i}\)\\
\midrule
0 &  2.50662827462874782e+00 & 1.00000000000000000e+00\\
1 &  7.07601239545194982e+00 & 3.32292052116718439e+00\\
2 &  7.79838748842165330e+00 & 4.35214422048696914e+00\\
3 &  4.28317532357895026e+00 & 2.86841498456426436e+00\\
4 &  1.24534893861795548e+00 & 1.01064393092251747e+00\\
5 &  1.88932466067008031e-01 & 1.87748433761637329e-01\\
6 &  1.38619741425921022e-02 & 1.70940757504186268e-02\\
7 &  4.15460128562968034e-04 & 6.51089901103861003e-04\\
8 &  3.33224453410284549e-06 & 7.12577601891527465e-06\\
9 & -1.65051672752885574e-09 & \\
\bottomrule
\end{tabular}
\end{center}

\lfktwentyc{} uses the same OTM coefficients as \lfktwenty{} in all three
shifted \(\tilde\eta\) zones.

\section{LFK-2026D: BigFloat Accuracy Optimised}\label{app:LFK2026D-coeffs}
The three OTM zones of LFK-2026D are
\[
\begin{array}{c|c}
j & \tilde\eta\text{ range}\\
\hline
1 & [0,0.0145)\\
2 & [0.0145,0.1178)\\
3 & [0.1178,1].
\end{array}
\]

The coefficients for this LFK-2026D variant are as follows.

\subsection{OTM branch, zone 1}

\begin{center}
\begin{tabular}{rNN}
\toprule
\(i\) & \(a^{26}_{1,i}\) & \(b^{26}_{1,i}\)\\
\midrule
0 & 1.24910559446641112e+00 & 1.00000000000000000e+00\\
1 & 8.47619772864081597e+02 & 9.64273267539114840e+02\\
2 & 2.99702508527653001e+05 & 4.43171035283801029e+05\\
3 & 6.53920879898315892e+07 & 1.24538153134200945e+08\\
4 & 9.45245193184535408e+09 & 2.31235285898333626e+10\\
5 & 8.92378101657744629e+11 & 2.87260538833032178e+12\\
6 & 5.19998272533330938e+13 & 2.30100251568817812e+14\\
7 & 1.57119179781390850e+15 & 1.06746011913009540e+16\\
8 & 1.78880731995660620e+16 & 2.20357803266821088e+17\\
9 & 3.63093569588198400e+16 & 1.29402991182225946e+18\\
10 & -2.96291760969892680e+16 & \\
\bottomrule
\end{tabular}
\end{center}

\subsection{OTM branch, zone 2}

\begin{center}
\begin{tabular}{rNN}
\toprule
\(i\) & \(a^{26}_{2,i}\) & \(b^{26}_{2,i}\)\\
\midrule
0 & 1.25070935316808463e+00 & 1.00000000000000000e+00\\
1 & 4.91931764982681784e+02 & 6.81403825183943468e+02\\
2 & 9.27513975150481565e+04 & 1.95271993660958280e+05\\
3 & 7.86397585558160208e+06 & 2.88054137623462491e+07\\
4 & 2.88765027905221641e+08 & 1.76543913139240003e+09\\
5 & 4.55199642305969429e+09 & 4.49113820497479324e+10\\
6 & 2.88783135940116310e+10 & 4.69178670074972595e+11\\
7 & 5.89013711536382675e+10 & 1.79968388446012061e+12\\
8 & 6.97512731591760159e+09 & 1.60734143390380566e+12\\
9 & -1.47094092991961823e+10 & -6.97033196207605225e+11\\
10 & 2.83876889443327570e+09 & \\
\bottomrule
\end{tabular}
\end{center}

\subsection{OTM branch, zone 3}

\begin{center}
\begin{tabular}{rNN}
\toprule
\(i\) & \(a^{26}_{3,i}\) & \(b^{26}_{3,i}\)\\
\midrule
0 & 1.61982027094435455e+00 & 1.00000000000000000e+00\\
1 & 1.66989132560739989e+02 & 6.11508302127448815e+02\\
2 & 4.77794828200244592e+03 & 3.47884126130892910e+04\\
3 & 5.31817138355287098e+04 & 6.32535742841342231e+05\\
4 & 2.61260745170476090e+05 & 4.72538757008512411e+06\\
5 & 5.91063965520491474e+05 & 1.58891631710913163e+07\\
6 & 6.06698667324509122e+05 & 2.44793192915989943e+07\\
7 & 2.60718058985714306e+05 & 1.65611815097258985e+07\\
8 & 3.82246440507954903e+04 & 4.32860404065594450e+06\\
9 & 9.69726676635529088e+02 & 3.10428817652220372e+05\\
10 & -9.78571143231221896e+00 & \\
\bottomrule
\end{tabular}
\end{center}

\acknowledgments{This paper would not exist without the initial idea from Gary Kennedy.}
\conflictsofinterest{The author declares no conflicts of interest.}

\begin{adjustwidth}{-\extralength}{0cm}

\reftitle{References}

\end{adjustwidth}

\end{document}